\documentclass{article}
\usepackage{graphicx}

\title{\bf Motion of a relativistic particle and the
vacuum\footnote{{\it Physics Essays}, {\bf 10}, no. 3, pp. 407-416
(1997)}}

\author{{\bf Volodymyr Krasnoholovets} \\
 {} \\
Institute of Physics, \\ National Academy of Sciences, Prospect
Nauky 46,  \\ UA-03650 Ky\"{\i}v 39, Ukraine}

\date{Received 24 May 1994}

\begin{document}
\maketitle

\begin{abstract}

 A vacuum medium model is advanced. The motion of a relativistic
particle in relation to its interaction with the medium is
discussed. It is predicted that elementary excitations of the
vacuum, called "inertons," should exist. The equations of the
particle path in Euclidean space are derived. The motion is marked
by the basic quantum mechanical relations $E=h\nu$ and $Mv_0 = h/
\lambda$ (here, $\lambda$ is the amplitude of spatial oscillations
of the particle  along  the trajectory, i.e., the interval at
which the velocity of the particle is periodically altered from
$v_0$ to 0 and then from 0 to $v_0$; \ $\nu$ is the frequency of
these oscillations). Analysis is performed on the transition to
wave mechanics where $\lambda$ manifests itself as the de Broglie
wavelength, and $\nu$ is the distinctive frequency of the
"particle-wave".  A prerequisite for the wave solution to be
Lorentz-invariant is treated. A hypothesis for a plausible
hydrodynamic description of the relativistic particle motion is
covered.   \\

 \vfill
\end{abstract}

\vspace{5mm}

\section{Introduction}
\hspace*{\parindent}
     Orthodox nonrelativistic quantum mechanics is well known$^{(1)}$
to suffer from long-range action. Furthermore, nonrelativistic
quantum mechanics is not Lorentz-invariant, although checking any
theory for Lorentz-invariance is now considered to be the best
guarantee of its reliability. Relativistic quantum mechanics does
not suffer from the above difficulties. However, both these
theories work very well in the area of microscopic phenomena, and
experiment confirms this fact perfectly. In connection with this
it can be assumed that a hidden mechanism uniting these two
limiting cases should exist, and it may result in the universal
quantum theory.

   There are different ways of devising a solution to the problem.
One way is based on the causal interpretation of the particle
behaviour in the system. Causal (deterministic) concepts have been
developed now in many papers (see, e.g., Refs. 2 to 6), yet
modeling of possible structural peculiarities of the vacuum medium
in which the particle moves is not  presented in any of them. And
some papers treat the vacuum space in a more or less explicit form
as a discrete lattice$^{(7-11)},$ as a peculiar kind of a quantum
crystal,$^{(12)}$ or as a model of Planck aether$^{(13)}$.

In the preceding paper$^{(14)}$ we have postulated a specific
cellular structure  for  the physical vacuum and discussed
deterministic dynamics of a nonrelativistic particle. In Ref. 14
an introduction of the potential of the particle's interaction
with the vacuum showed that the motion of the particle is of an
oscillating character:  initially, at the section $\lambda/2$ of
the path, the vacuum medium knocks elementary excitations
(inertons) out of the moving particle. They migrate ahead of the
particle, and the velocity of the particle gradually decays to
zero and, subsequently, at the next section $\lambda /2$ of the
path, the particle  absorbs inertons, this time from behind, and
the quantity of  its velocity regains its original value $v_0$. As
this takes place, the amplitude $\lambda$ of the particle's space
oscillations is identified as the de Broglie wavelength.

     The present paper develops the model described in Ref. 14
when the relativistic spinless particle is in motion.

\vspace{4mm}

\section{Vacuum medium}
\hspace*{\parindent}
     The fundamental notions of modern geometry (in coordinate
approach) are  as follows$ ^{(15)}$: \ (1)  the point in space, \
(2) the vector of this point's motion velocity, \ (2) the line
length along which the  point  moves.  Some spaces (Euclidean,
Riemannian, and others) are specified above all by their metric.
In addition, we introduce  the discreteness of space, dividing it
into uniform cells the size of which is close to the value
$10^{-28}$ cm, as required by the grand unification of
interactions (all types of interaction come together in this
scale). Let us assume that each of these these cells contains a
superparticle, which is in the degenerate state over all possible
multiplets, since for  each elementary particle there is a certain
multiplet state. The degeneration of the superparticle may be
released spontaneously or in response to the external factor as a
result of which its volume changes and a certain symmetry appears,
that is,  an  elementary particle is created. In the author's
view, a change in volume  of the initial superparticle is
associated with the emergence of  the mass of the particle at
rest. Therefore, the particle  represents a defect in the
homogeneous discrete vacuum space; hence the space around this
defect must be rendered somewhat deformed: the neighboring cells
are subjected to deformation that may be identified  with the
gravitational potential appearing in the space surrounding the
particle.

  The motion of the physical "point" (particle cell) in  the entirely
packed discrete space is accompanied by the interaction with the
"points" of the space (superparticles cells), giving rise to
excitations in neighboring superparticles.  As  the  particle
moves, it constantly emits and absorbs excitations (inertons), but
in distinction to the real motion of the particle, the motion of
inertons is different: they migrate by a  relay  mechanism, from
superparticle to superparticle. Since inertons are elementary
excitations of the vacuum medium, their velocity, at least the
initial velocity, is of the order or equal to the speed of light
$c$, the velocity of the particle $v_0<c$. Let us consider the
motion of such system in the relativistic approximation.

\vspace{4mm}
\section{Lagrangian and the particle trajectory}
\hspace*{\parindent}
     In Ref. 14 we proceeded from the  modernized classical
nonrelativistic Lagrangian of the particle of the type
 $$
L_{\rm nonrel.}={ 1\over 2} g_{ij} {\dot X}^i {\dot X}^j + U(X, x,
\dot x), \eqno (1)
 $$
where the function $U$ accounted for the potential energy of the
interaction between the particle and inertons and the kinetic
energy of inertons (hereinafter the vectors $X$ and $\dot X$
pertain to the particle, the vectors $x$ and $\dot x$ pertain to
inertons). On examination of the relativistic particle, we shall
proceed in a similar fashion, that is, in the classical
relativistic Lagrangian
 $$
L_{\rm rel.} = - M_0 c^2 \sqrt{1-v_0^2/c^2}  \eqno (2)
 $$
we shall substitute
  $$ v_0^2 {\longrightarrow  { [g_{ij}\dot X^i
\dot X^j +U(X,x,\dot x)]/g}}.
    \eqno (3)
  $$

    The modernized relativistic Lagrangian of the  particle,  in
view of the interaction with inertons, is written in an  explicit
form as (cf. with  Ref. 14):
 $$
 {\cal L} = -gc^2 \left\{ 1 - {1\over gc^2} \left[
g_{ij}\dot X^i (t)\dot X^j(t) +\sum_{r=0}^{N-1}\widetilde
{g}^{(r)}_{ij}\dot x^i_{(r)}(t_{(r)})\dot x^j_{(r)}(t_{(r)})
\right. \right.
  $$
  $$-\sum_{r=0}^{N-1} {2 \pi\over T_{(r)}}\delta _{t-\Delta t_{(r)},
   t_{(r)}} \left(X^i(t)
   \sqrt
{g_{is}(\widehat {A}^{-1}\widetilde {g}^{(r)}_{sj})_0} \ \  \dot
x^j_{(r)}(t_{(r)})\right.
  $$
  $$ +\left.\left.\left.\dot
X^i(t)\Big\vert_{t=0}\sqrt{ g_{is}(\widehat {A}^{-1}\widetilde
{g}^{(r)}_{sj})_0} \ \  x^j_{(r)}(t_{(r)})\right) \right ] \right
\}^{1/2}. \eqno(4)
  $$
Here, $g_{ij}$ are components of the metric  tensor produced by
the particle in the three-dimensional space; along the trajectory
of the particle\  $g_{ij} = {\rm const} \ \delta _{ij}$ \ and
$g=g_{ij}\delta^{ij}$ is the convolution of the tensor;
$\widetilde {g}^{(r)}_{ij} (x_{(r)})$  are components of the
metric tensor of the {\it r}th inerton in the three-dimensional
space, along the trajectory of the {\it r}th inerton the tensor
$\widetilde {g}^{(r)}_{ij} $ is supposed to be locally equal to \
${\rm const} \ \delta _{ij}$   \   (index {\it r}  is enclosed in
parentheses to distinguish it from  indices {\it i, j} and {\it s}
that describe  tensor and vector quantities); $1/T_{(r)}$ is the
frequency of collisions of the particle with the {\it r}th
inerton, $N$  is the number of inertons emitted by the particle;
$\delta _{t-\Delta t_{(r)}, t_{(r)}}$ is Kroneker's symbol that
provides the agreement between the proper time $t$ of the particle
and $t_{(r)}$ that of the {\it r}th inerton  at the instant of
their collision ($\Delta t_{(r)}$ is  the time interval after
expiry of which, measuring from the  initial moment $t  = 0$, the
moving particle emits an inerton). The operator $\hat A^{-1}$
allows for  the motion of the metric in every inerton at the
instant of its emission (absorption) by the particle: $\widetilde
 {g}^{(r)}_{sj}\rightarrow \widetilde  {g} ^{(r)}_{s+u, j}$ (with
regard to  cyclic permutation,  indices {\it s,  j}    and {\it u}
take on  values 1, 2 and 3), that is, it shifts the emitted
inerton to the trajectory distinct from the path of the particle.
As the proper time {\it t}   of the particle is expressed through
the proper time $t_{(r)}$ of  the {\it r}th excitation,
$t=t_{(r)}+ \Delta t_{(r)}$ (see Ref. 14), it suffices to write
Euler-Lagrange equations for the particle and the {\it r}th
inerton in terms of one of these two time parameters, for
instance, via $t_{(r)}$: $$ d(\partial {\cal L}/\partial {\dot
Q^k})/dt_{(r)} - \partial {\cal L}/\partial Q^k = 0 \eqno (5) $$
where for the particle $$ Q^k \equiv X^k (t_{(r)}+\Delta t_{(r)})
\eqno (5a) $$ and for the {\it r}th inerton $$ Q^k \equiv
x^k_{(r)}(t_{(r)}). \eqno (5b) $$

     When moving, the particle constantly  exchanges  the  energy
and momentum with the ensemble of inertons.  In  this  case,  the
laws of conservation of the total energy $E$  and the total
momentum  $p_0$  for the particle-inertons' ensemble system should
be  obeyed along the path {\it l}. The constants $E$   and
 $p_0$   are obviously the initial values for the particle at the
trajectory [at  initial instant {\it t} = 0, the Lagrangian (4)
still retains its  classical  form: expression (2), in terms of
which quantities $E$   and  $p_0$   are derived using the standard
 method; see, e.g., Refs. 16, 17]. Since the  expression under the radical
symbol in the Lagrangian (4) is  constant,  we may treat the time
$t$ as the parameter proportional to the natural  {\it l} (
evidence for this statement, for  instance,  for the Lagrangian
$L=\sqrt {g_{ij}\dot X^i \dot X^j}= {\rm const}$, see  Ref. 18).
 However, as time is considered to be the
parameter accordant with the particle path length, $t  \propto l$
(for inertons $t_{(r)}\propto l_{(r)}$,  where $l_{(r)}$ is  path
length for  {\it r}th inerton),  equations  of  extremals for the
Lagrangian (4) are in complete agreement with the  appropriate
equations obtained in  Ref. 14 for the nonrelativistic case (see
Ref. 19,  which shows identity  of  equations for extremals
derived from the Lagrangians  $L = g_{ij}\dot X^i \dot X^j$ and
$L=\sqrt { g_{ij}\dot X^i \dot X^j}$, with the allowance made for
a natural character  of the  parameter $t$).

     Some details for solving equations of motion for the particle and
inertons are adduced in Appendix. We shall present here the
ultimate  result true to one period $T_{(r)}$ of collisions, that
is, the time interval between  the moments of emission and
absorption of the {\it r}th inerton by the particle. If the
particle moves along the axis $X^1 \equiv X$, then the motion of
inertons is defined by the coordinates longitudinal and
transversal to the $X$-axis, that is,
 $x^{\parallel}_{(r)}$ \  and  \
$x^{\perp}_{(r)}=[(x^2_{(r)})^2 + (x^3_{(r)})^2]^{1/2}$, \
respectively (hereinafter, we omit the parentheses for indices
$r$):
  $$
\dot x^{\perp}_r (t_r)= c \cos (\pi t_r/ T_r); \eqno (6)
  $$
  $$
x^{\perp}_r(t_r)= {\Lambda \over \pi} \sin (\pi t_r/ T_r); \eqno
(7)
  $$
$$ \Lambda _ r = c T_ r. \eqno (8)  $$
For the longitudinal coordinate
  $$
   \dot x^{\parallel}_r = {\rm const} =3v_{0r}/{ 2 \pi} \eqno (9)
  $$
where $$ v_{0r} = v_0  [1 - \sin (r\pi / 2N)], \eqno (10) $$ $v_0$
is the initial velocity of the particle at the moment $t_r=t= 0$.
In formulas (6) to (9), the variable $t_r$   determines the proper
time of the { \it r}th inerton within the time interval  $T_r$.
For the particle the solutions take the form
 $$
\dot X(t_r)=v_{0r}[1 - \sin (\pi t_r/ T_r)]; \eqno (11)
 $$
 $$
X(t_r)=v_{0r}t_r +{\lambda _r \over \pi}[\cos (\pi t_r/ T_r) - 1];
\eqno (12)
  $$
  $$ \lambda _r = v_{0r}T_r. \eqno (13) $$
As seen above, the solutions (11) and (12) are written in terms of
the proper time of the {\it r}th inerton. In Ref. 14 solutions
(6), (7), (9), (11) and (12) were lengthened  for the next
half-periods of oscillations $nT_r$, where $n$  is the number  of
the spatial oscillation of the particle along the trajectory from
the initial point of its motion.  However,  the quasicontinuous
time parameter $t_{rn}$   has been introduced for this  purpose.
In  the case of the relativistic Lagrangian (4), the parameter $t$
is regarded as being natural, proportional to  the trajectory
length {\it l}; hence this parameter should necessarily be
continuous. This requirement may be readily satisfied, if to the
initial condition $\dot X\mid _{t=0}=v_0$ [see formula (A11)] we
add another requirement in accordance with which the quantity of
the particle velocity for the {\it n}th time period of collisions
$T$ resumes its initial value, that is,
  $$
\dot X\big\vert_{t=0} = \dot X\big\vert_{t=nT}= v_0,\ \ \ n = 1, \
2, \ 3, ... \eqno (14)
  $$
[with  arbitrary $t$,  variable $\dot X (t)\geq 0$].  At this
condition the solutions from (11) to (13) as functions of the
proper  time $t$ of the particle   assume the form (by the way,
true and, besides, more attractive one for  nonrelativistic case):
  $$ \dot X (t )= v_0 (1 - \big\vert \sin (\pi t /T) \big\vert );
\eqno (15 a)
  $$
  $$
X(t) =v_0t + {\lambda\over\pi} \Bigl\{ (-1)^{[t/T]} \cos(\pi t/T)
-\bigl( 1+
 2[t/T] \bigr) \Bigr\}; \eqno (15 b)
  $$
  $$ \lambda = v_0 T.  \eqno (16)
  $$
Equation (11) has been integrated over each {\it n}th interval, so
the solutions for all $n$ intervals have been united. In this case
$X$ as a function of $t$ in $(15b)$ is the continuous and positive
value with the initial condition $X(0)=0$. The notation $[t/T]$ in
$(15b)$ means an integral part of the integer $t/T$.

        Motion trajectories of the particle and surrounding inertons
are illustrated schematically in Fig. 1.
\begin{figure}
\begin{center}
\includegraphics[scale=0.7]{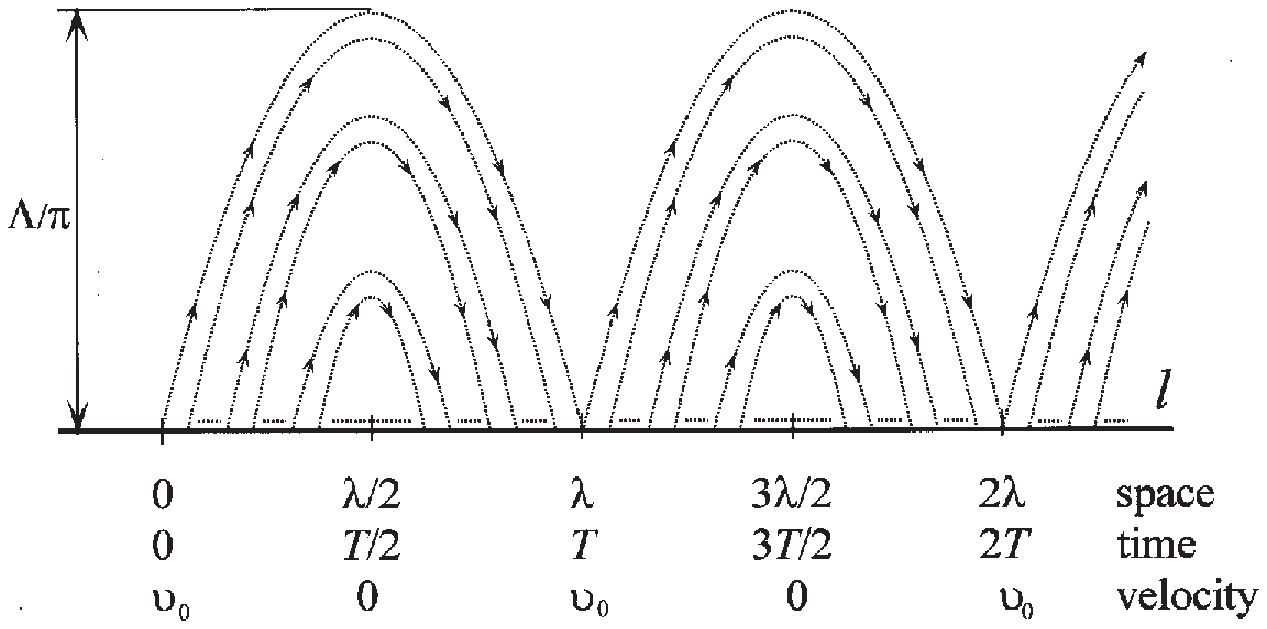}
\caption{The schematic representation of the motion trajectories
of the particle (along its path $l$) and attending inertons.}
\label{Figure 1}
\end{center}
\end{figure}
As can be seen, the path of the particle may be divided into equal
sections of  length $\lambda$, even though the parameter $t$  is
continuous along  the trajectory.  The parameter $\lambda$  should
be regarded as the space period or as  the amplitude of the
particle's spatial oscillations. It is of interest that in
conformity with correlation (16), the time half-period of
oscillations $T$  may be written as $T= \lambda /v_0$ (for
inertons $T_r = \Lambda_ r / c$ ). Thus the quantity $T$ is the
natural parameter  at  the length of the spatial period $\lambda$
for the particle, much as its proper time $t$  is the natural
parameter for  the  entire  trajectory, $t = l / v_0$. The
quantity $T_r$  may be viewed as the natural  parameter at the
length of the spatial period $\Lambda_r$   of the {\it r}th
inerton. However, for each portion the proportionality
coefficients between $T$  and $\lambda$   for the particle as well
as $T_r$ and $\Lambda_r$ for inertons differ essentially and are
equal to $1/v_0$ and $1 / c $,  respectively.

\vspace{4mm}
\section {Wave relativistic mechanics}
\vspace{2mm}
\hspace*{\parindent}
     If the ensemble of inertons is considered  as  the whole
object, an inerton cloud with the effective mass at rest $m_o$,
then the original Lagrangian should be accordingly altered. In
Euclidean space it may be presented as
 $$
L =  - M_0 c^2
   \Bigl\{ 1  - {1\over M_0 c^2 }
 \Bigl[ M_0 \dot X^2 + m_0 {\dot x}^2
  $$
  $$-\frac{2\pi}{T}\sqrt {m_0 M_0}\ (X \dot x+ v_0 x) \Bigr]
\Bigr\} ^{ 1/ 2}.   \eqno (17)
 $$
The function (17) describes the particle with the mass at rest
$M_0$ that moves along the $X$-axis with the velocity $\dot X$
($v_0$  is  initial velocity); $x$  is the distance between the
inerton cloud and the particle, \  $\dot x$   is the  velocity of
cloud in the frame of reference connected with the particle, and
$1/T$ is the frequency of collisions between the particle and
inerton cloud. We believe that $x$ and $\dot x$ are functions of
the proper time $t$ of the particle.

     Euler-Lagrange equations for the Lagrangian (17) are reduced to the
following system [for this purpose relationship
$M_0/m_0=c^2/v_0^2$ from formula (A6) is used]
  $$
\ddot X +{\pi \over T}{v_0\over c}\dot x = 0; \eqno (18)
  $$
  $$
\ddot x -{\pi\over T}{c\over v_0}(\dot X - v_0)=0. \eqno (19)
  $$
The solutions $x(t)$ and $\dot x (t)$ obtained from Eqs.(18), (19)
have the form $$
    x ={\Lambda \over \pi}|\sin (\pi t / T)|, \eqno (20a)
$$
 $$
\dot x = c(-1)^{[t/T]}\cos (\pi t / T) \eqno (20b)
$$
where $\Lambda$ is the amplitude of the inerton cloud oscillation.
The solutions $X(t)$ and $\dot X (t)$ comply with expressions (15).
Solutions of $X, \ \dot X$ and $x, \ \dot x$ are shown in Fig. 2.
Notice that for the decision of Eqs. (18), (19) initial conditions
$x(0)=X(0)=0$, condition (14), inequalities $x, X, \dot X \geq 0$
and $c\geq \dot x \geq -c$ have been taken into account.
\begin{figure}
\begin{center}
\includegraphics[scale=0.8]{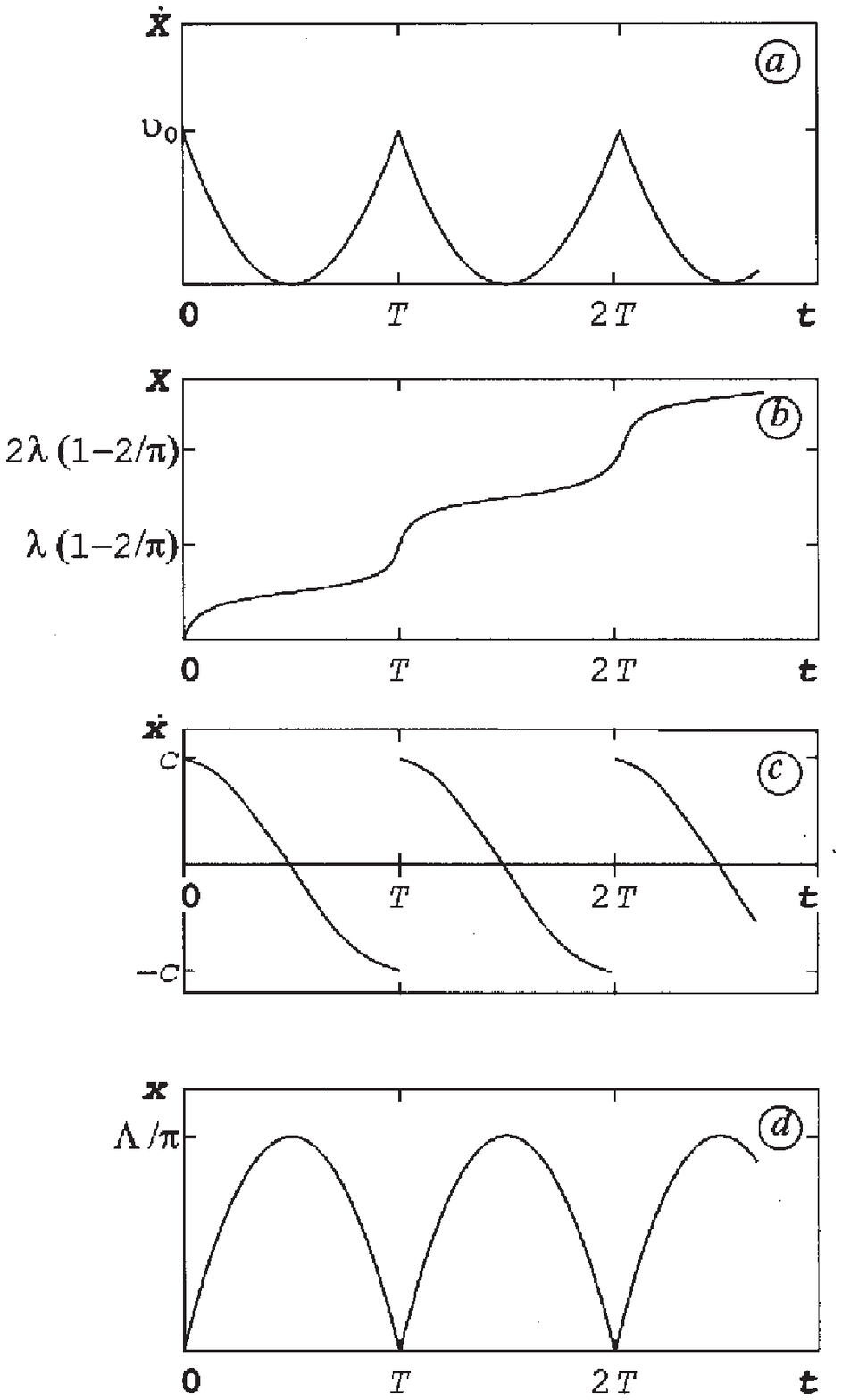}
\caption{The schematic representation of the solutions of
         Eqs. (15) for the particle  $(a, b)$ and Eqs. (20)
         for the inerton cloud $(c, d)$.
} \label{Figure 2}
\end{center}
\end{figure}

     While substituting
$$
{\dot{\widetilde x}}= \dot x - {\pi \over T}\sqrt {M_0/m_0}\ X
\eqno (21)
$$
the Lagrangian (17) is reduced to the canonical form
on variables for the particle:
  $$
L =  - M_0 c^2 \Bigl\{ 1 - {1\over M_0  c^2 } \Bigl[
    M_0 \dot X^2 - M_0 (2\pi/ 2T)^2 X^2   \ \ \ \ \ \ \ \ \ \ \ \ \ \
    \ \ \ \ \ \ \ \
 $$
  $$\ \ \ \ \ \ \ \ \ \ \ \ \ \ \ \ \ \ \ \ \ \ \ \ \ \ \ \ \ \ \ \ \ \ \ \
    + m_0 \dot {\widetilde x}^2 - \frac{2\pi}{T} \sqrt {m_0 M_0}\ v_0
    x \Bigr] \Bigr\}^{ 1/2}.   \eqno (22)
  $$
As seen above, the first two terms in brackets under the
radical in (22) describe the particle alone and comply with the
kinetic and potential energies of the harmonic oscillator; the
last  two terms depict the effective kinetic and potential
energies of the inerton cloud.

     Let us introduce  the Hamiltonian   function  conforming to the
Lagrangian  (22). In accordance with the definition
       $$H = \sum _i \dot Q_i \partial L / {\partial{\dot Q}_i} - L. $$
 Therefore, for our case we have
 $$ H = \dot X \partial L / {\partial \dot X} + \dot {\widetilde x}
{\partial L}/{\partial \dot {\widetilde x}}   - L; \eqno (23)
  $$
hence for the function $L$ (22) we derive the corresponding
Hamiltonian
  $$ H=M(2\pi / 2T)^2X^2+Mc^2+\pi\sqrt{m M_0} \ v_0 x/T
\eqno(24)
  $$
where
 $$
  M = M_0 /\sqrt{1-v_0^2/c^2},\ \ \ \ \ \   m=m_0/\sqrt{1-v_0^2/c^2}.
 $$
When deriving  (24), we have paid attention to the fact that in
the Lagrangian  (22)  the  radical $\sqrt{...}$ \ is const, equal
to $\sqrt{1-v_0^2/c^2}$. On the other hand, in compliance with the
derivation of momenta for the particle $p$ and of the inerton
cloud $\widetilde p$,  we obtain
  $$ p =
\partial L /
\partial {\dot X} = M \dot X; \eqno (25) $$ $$ {\widetilde p} =
\partial L /\partial{\dot {\widetilde x}} = m {\dot {\widetilde
x}}. \eqno (26)
  $$
Thus with allowance made for the expressions (25), (26) and (22),
the Hamiltonian (23) may be presented as
  $$ H=p^2 / M  +
{\widetilde p \ }^2 / m + (M_0 c)^2 /M. \eqno(27)
  $$
Both presentations (24) and (27) of the relativistic Hamiltonian
are equivalent. By combining (24) and  (27), the following
Hamiltonian is produced
   $$
   H=p^2/2M  +  M(2\pi/ 2T)^2 X^2 /2
    +  [(Mc)^2+(M_0 c)^2]/2M
  $$
 $$ +{\widetilde p}^2/2m + \pi\sqrt{m M_0}
    v_0 x/ T. \eqno(28)
$$ The first two terms in (28) describe the particle  and
represent the Hamiltonian of the harmonic oscillator; the third
term is the renormalized energy of the particle at  rest,  and
the last  two terms are the kinetic and renormalized potential
energies of  the inerton cloud.

     Let us separate out of (28) the effective Hamiltonian of the
particle that describes its behavior relative to the inertia
centre of the particle-inerton cloud system:
  $$
H_{\rm eff}=p^2/2M+M\left(2\pi/ 2T\right)^2X^2/2. \eqno (29)
  $$
As done in Ref. 14, we now turn our attention  to the function of
the action. For a shortened action $S_1$ of the particle we have
the Hamilton-Jacobi's equation:
  $$
{\left(
\partial S_1/\partial X \right)}^2/2M + M\left(2\pi/
2T\right)^2X^2/2     = E. \eqno (30)
  $$
For the action  $S_1$ from Eq. (30) we acquire
  $$ S_1 = \int^X p
dX = {\int} ^ X \sqrt {2M [ E- M\left(2\pi/ 2T\right)^2X^2/2 ]}\
dX. \eqno (31)
  $$

   The solution for $X$ as the function of $t$ has the
appearance (see, e.g.,  Refs. 20, 21)
 $$
X=\sqrt {2E\over M(2\pi/2T)^2 }\ \sin (2\pi t/2T). \eqno (32)
 $$
The periodicity in the particle motion allows to pass on  to the
action-angle variables in Eq. (31) that depicts this  motion.
Hence for the increment of the particle action within the cyclic
period $2T$   we obtain
      $$J=\oint p dX = E\times 2T. \eqno (33)$$
Allowing for an agreement between the  constant $J$ and  the
initial kinetic energy of the particle $E=Mv_0^2/2$, expression
(33)  may be rewritten as
      $$J=Mv_0 \times v_0 T = p_0\lambda\eqno (34)$$
where $p_o$ is the initial momentum, and, when  deriving (34), the
relation (16) is used. By assuming that  $J=h$  ( $h$ is the
Planck constant), we obtain the de Broglie relation (for
relativistic particle) from (34):
        $$h/\lambda=p_0\equiv M v_0. \eqno (35)$$
According to our interpretation, in formula (35) $\lambda$ is the
spatial period of the particle oscillations on the trajectory.
Since $2T$ is the cyclic period of the particle oscillations, then
by introducing the appropriate frequency $\nu = 1/2T$ into (33)
and assuming that $J=h$, we obtain another of the two main
relations of quantum mechanics:
         $$E=h\nu. \eqno (36)$$
But it should be noted that the constant  $E$   is consistent with
the initial  kinetic  energy  of  the   relativistic   particle,
$E= Mv_0^2/2$.

     In configurational space where the solution  (32)  for  $X$
is found, the variable $X$   is not restricted by the spatial
period $\lambda$.  Here, the trajectory of the particle is
continuous and in the course of time the particle moves steadily
away from the initial point $X=0$. Relations (35) and (36) admit a
correlation between the particle that constantly moves along the
$X$-axis of configurational space and a plane monochromatic wave
that travels in the same direction and is marked by the frequency
$E/h$ and wavelength \  $h/p_0$. This correlation is
known$^{(22)}$ (see also Ref. 14) to result in Schr\"odinger wave
equation
  $$ (\hbar ^2/2M) {\nabla}^2\psi +  E\psi =0   \eqno(37)
  $$
where the wave function
  $$ \psi = {\psi}_0 \exp[2\pi i (X/\lambda
- \nu t)].   \eqno (38)
  $$

\vspace{4mm}
\section  {Discussion}
\hspace*{\parindent}
     Schr\"odinger equation (37) is somewhat specific as it describes
 the behavior of the particle with the relativistic mass $M=
M_0/\sqrt{1-v_0^2/c^2}$.  In this equation, the momentum operator
\ \  $-i\hbar \nabla $    and the eigenvalue  $E$  are related to
the relativistic  momentum  $Mv_0$ and relativistic kinetic energy
in the form of $Mv_0^2/2$, respectively. In the approximation
$v_0^2/c^2\ll 1$, Eq. (37) reduces to a common nonrelativistic
Schr\"odinger equation. Simply a part of the particle's  total
energy, namely, its kinetic component, $Mv_0^2/2$ appears in Eq.
(37) due to the obtained correlation (36), where precisely the
kinetic energy of the particle-inerton cloud system defines the
frequency of the system's oscillations.

    The equality  $E=h\nu$ is formally postulated for the
particle in quantum  mechanics;  both of the quantities $E$ and
$\nu$  are indeterminate in this correlation: $E$ at times imply
just the kinetic energy $M_0 v_0^2/2$ of the nonrelativistic
particle (see, e.g., ${\rm Fermi^{(23)}}$ and ${\rm Schiff
^{(24)}}$) and at other times the total energy $Mc^2$ (see, e.g.,
${\rm Schiff^{(25)}}$ and  ${\rm Sokolov\ et\ al. ^{(26)}}$). De
Broglie himself, having introduced the  relation (36)  for the
particle, adhered to both of these contrary opinions relative to
$E$  at different times (see Ref. 27 and comments by Lochak in
it). But the fact that the proper frequency $\nu$ attributable to
the moving particle cannot be determined  by  its total  energy
$Mc^2$ may be supported by the argument that precisely this
formula is applied for description of annihilation process. For
instance, by annihilation of an electron and positron that have
the same energies  $E_{\rm tot}\equiv Mc^2$, the energy $h\nu$ of
the two emitted photons is found precisely from the equation
$2E_{\rm tot}=2h\nu$. Therefore, it is not logical to ascribe the
frequency $\nu =E_{\rm tot}/h$ to the particle as this correlation
automatically defines the particle instability in relation to its
transmutation. At the same time, the above-obtained relation (36),
where $E$ is simply the kinetic component $Mv_0^2/2$ of the total
energy $Mc^2$, is internally noncontradictory in the context of
the proposed model.

     We shall now highlight another important detail  related  to
Eq. (37) or, more exactly, to the generalized wave equation
  $$
(i\hbar {\partial \over \partial t}+{{\hbar}^2\over
2M}{\nabla}^2)\psi=0 \eqno(39)
 $$
where the $\psi$-function is defined in (38). As mentioned above,
time $t$  is the proper time of the particle proportional to the
natural parameter $l$; namely, in the case of the Lagrangians (4)
and (17) the proper time $t$ has been chosen in the form of
$t=l/v_0$. In other words, we suppose that the form has been
predetermined by the initial value of the particle velocity $v_0$.
So the chosen proper time $t$ is the proper time along the world
line $l$, and, according to definition (see, e.g. Refs. 28 and 29)
these two parameters ($l$ and $t$) should be invariant Lorentz
transformations. The path is made  up  of identical sections of
length $\lambda$; hence the invariance $\lambda$ follows from
Lorentz invariance of the quantity $l$, and since the time
half-period $T$ of the cycle  of the particle is proportional to
$\lambda$, that is, $T=\lambda/v_0$, then the quantity $T$ will
also be invariant. Thus the variables $X$ and $t$ shared by the
spatial $\lambda$   and time $T$  intervals, respectively, are
Lorentz-invariant in the wave equation (39).

     Each interaction geometrization stage in physics calls for a
penetrating insight into physical processes of the phenomenon  we
attempt to formalize. Thus in solids and liquids  relativistic
effects apparently stem from a change in intermolecular and
interatomic forces, and it is just this hidden  dynamics  that
should be the basis for the formalism of  special  relativity.

  Lorentz also indicated this  possibility$^{(30,31)}$. In 1902
and 1904 he proposed hypotheses on the change due to the motion of
the intermolecular forces, geometry of the electron, and its mass
in the aether. In 1913 Ehrenfest$^{(32)}$ analyzed views of
Lorentz and Einstein (in 1905 the latter suggested$^{(33)}$ that
the speed of light in a vacuum is constant). Ehrenfest came to the
conclusion that Einstein's special relativity, which denied the
aether, leads to the same results as Lorentz's aether theory. In
this case, he had stressed, such {\it experimentum crucis} which
could solve the controversial question in favor of this or that
theory cannot be realized. Nevertheless, as a consequence of the
postulate of the universality of the speed of light $c$ and the
principle of equivalence of an infinite multitude of inertial
frames of reference forced out the notion of the orthodox aether.
So the use of Lorentz kinematic transformations have been
sufficient to give the final results.

These transformations are universal for all measured physical
magnitudes (coordinate, time, electric and magnetic fields, etc.).
Therefore, it really seemed quite natural to drop the notion of an
aether medium with its inner deformations which are not measurable
in principle. Why should this medium be introduced into the space
between material objects, if any instrument by its nature is
oriented only to a measurement of characteristics connected solely
with those material objects? It was just special relativity that
offered the clue to finding the change in physical magnitudes when
an object moves, and a notion of an intermediate medium between
objects is not necessary for this purpose.

     This is undoubtedly so. However, on the other hand, all the
modern significant achievements in the solid area are based on the
noncausal and nonrelativistic, without Lorentz-invariance quantum
mechanics. Thus for decades we found ourselves in a paradoxical
situation because we must trust both the special theory of
relativity, which has been substantiated by numerous experiments,
and the nonrelativistic theory (quantum theory), which constitutes
the foundation of all solid-state science. How does one resolve
this paradox? Logic clearly indicates that it is necessary to turn
to the notion of actual aether medium but taking into account
contemporary knowledge. (Spectacular proof is provided by the
physics of elementary particles in which various bag models,
bubble models, lattice models, string models, etc., are paramount
important. It is very difficult to consider these subquantum media
merely as implicit forms.)

    In the proposed theory (this paper and Ref. 14)
the notion of a discrete vacuum medium is original. Hence this
medium allows the creation of the the causal dynamics of an
elementary particle and obtaining the equation of a particle path
(15). However, there are the problem of conforming the
peculiarities of the vacuum medium with special relativity. It
should be noted that we already proceeded from the relativistic
Lagrangian invariant (2), and it allowed us to obtain the
Schr\"odinger equation in the invariant form (39). Furthermore,
the reduction in size of the particle and superparticles as well
as the availability of the adiabatic motion of the former are also
related to the peculiarities of the medium.

       By  definition,$^{(14)}$ the mass of the particle as a local
curvature in the vacuum medium is  the  ratio  between the
superparticle volume {\it V}  in the degenerate, that is, in the
nondeformed, vacuum and the particle  volume $V_0$, that is,
  $M_0={\rm const} \ V/V_0$.
As the particle moves along the $X$-axis, its reduction in size by
a factor of $\sqrt{1-v_0^2/c^2}$  \   in this direction
automatically leads to an increase in its mass by a factor of $(1
-v_0^2/c^2 )^{-1/2}$; \ \ \ indeed, as\ \  $M_0\propto V^{-1}_0=
(R_{0x}R_{0y}R_{0z})^{-1}$,\ \ then $$ M=M_0/\sqrt{1-v_0^2/c^2} \
\propto\  1/[(R_{0x}\sqrt{1-v_0^2/c^2} \ )R_{0y}R_{0z}] \eqno(40)
$$ where $R_{0i}$  is the typical size of the particle in the
state of rest along the axes $i= \{X, Y, Z \}$.

     Nonetheless, the result (40) may be obtained from purely  physical
consideration, namely, since the motion of the particle together
with the surrounding deformation coat is regarded as the travel of
an element of a liquid in hydrodynamics; the appropriate equation
takes the   form$^{(34)}$ $$ \rho\  d\vec{v}/dt =- \nabla {\cal
P}\eqno (41) $$ where $\rho$   is the liquid element density and
$\cal P$ is the pressure of the liquid on the moving element. We
assume that the motion  is adiabatic, that is, the change in
pressure on the side of the liquid upon the element chosen is
proportional to the  variation in density of this element and
then$^{(35)}$ $$ (\partial{\cal P}/ {\partial {\rho}})_{\rm
entropy} =c^2 \eqno(42) $$ where {\it c} is the maximum velocity
for this liquid (sound velocity). Equation (41) is nonlinear and
consequently allows for multiple solutions. However, there is only
one possibility when this equation becomes linear, and, therefore,
there is a single solution. That situation is realized for our
medium where the motion is highly particular.

     In hydrodynamics, the point is limited by the proper size of the
element of the liquid. This size is enormous as compared with the
particle size. Hence the character of change of the substantial
derivative $d{\vec v}/dt$ in Eq. (41) should be specified by
microscopic processes which take place in the vacuum medium when
the particle moves. The main peculiarity that results from these
processes is nonstationary motion of the element of our liquid:
from the initial moment of the motion its velocity changes from
$v_0$ to 0, then from 0 to $v_0$, etc., with each time interval
$T/2$. Thus a tangible change in the velocity of the element is
observed over the time interval $T/2$ and spatial interval
$\lambda/2$. The smallest scale does not appear at the continuous
consideration of the motion of that liquid's element. For this
case the substantial derivative may be defined as $$ {df(x)\over
dx}=\lim_{\Delta x \rightarrow \left\{
         \matrix{T/2\cr
         \lambda/2\cr} \right\}} {f(x+\Delta x) -f(x)\over {\Delta x}} =
         \lim_{\Delta x \rightarrow \left\{ \matrix{ T/2\cr
                                                  \lambda/2\cr} \right\}}
         {\Delta f\over \Delta x}.
         \eqno (43)
$$
In this connection, continual Eq. (41), with regard for (42),
necessitates a substitution for its discrete analogy:
$$
\rho\  \Delta v /  \Delta t =-c^2 \Delta \rho /
         \Delta l; \eqno (44)
$$ here,  $\Delta l = \lambda/2$ and $\Delta t=T/2$ are the
spatial and  time  intervals within which the element velocity
decreases from the initial value  of  $v_0$   to the final 0,\ \
that is, \ $\Delta v = -v_0$. \ When the speed of the element
changes from the maximum to the minimum magnitude ($v_0
\rightarrow 0 $), the pressure, on the contrary, changes from the
minimum  ($\cal P$) to the maximum  (${\cal P}_0$) magnitude;
therefore, $\Delta {\cal P}={\cal P}-{\cal P}_0$. Then, from the
formula $\Delta {\cal P}/\Delta \rho=c^2$ \ (42) one finds that
$\Delta \rho =\rho-\rho_0$ where $\rho$ and $\rho_0$ are the
density of the liquid element at the moment of motion and in the
state of rest, respectively. Hence, from (44),  with regard  to
the correlation $\lambda/T=v_0$, we gain
  $$
\rho = \rho_0/(1- v_0^2 /c^2). \eqno (45)
  $$
At the next stage of element motion the parameters of Eq. (44) are
as follows: $\Delta l=\lambda/2$,\ $\Delta t=T/2$, $\Delta v=v_0$,
and $\Delta\rho=\rho_0-\rho$; and again one obtains expression
(45). This expression is true even at $v_0 \rightarrow c$.

    Our liquid element is essentially the deformation coat that
surrounds the particle. Since the coat is a linear response of the
vacuum medium to the particle creation, then the total coat mass
should be equal to the particle mass, that is, to $M_0$ in the
state of rest. This is apparently  true for the volume $V_0$,
because, according to our hypothesis, the mass of the particle is
equivalent, with accuracy to the constant factor, to its inverse
proper volume. The value $V_0$ for the element plays the role of
an effective volume, but for an arbitrary element of the
degenerate vacuum (where there is no particle) that the effective
volume is absent, $V_0=0$.

      So, we have for the liquid element $M_0\propto 1/V_0$ and,
therefore, $\rho_0\propto 1/V_0^2 $. In agreement with (45), the
motion  of  the  element with the velocity $v_0$ results in a
decrease of the total volume of the element in the direction of
the motion: $\rho_0\rightarrow\  \rho \propto
1/(V_0\sqrt{1-v_0^2/c^2} )^2$.\ Then the  mass of the element and
the mass of the particle as well (due to the  interdependence of
these masses) change:
    $$
M_0 \propto 1/V_0 \rightarrow 1/(V_0 \sqrt{1-v_0^2/c^2}),
\eqno(46)
    $$
that is, the mass increases in the direction of the velocity
vector by a factor of $1/\sqrt{1-v_0^2/c^2}$    \   which is
likely to be in complete agreement with the experimental fact and
formalism of special relativity.

\vspace{4mm}
\section{Conclusion}
\hspace*{\parindent}
        The present concept is based on the availability of a discrete
vacuum medium. This medium (quantum aether) is not contradictory
to special relativity because the model:\  (1) satisfies
Einstein's postulate of the universal speed {\it c} in a vacuum,
that is, the postulate corresponds to the adiabatic nonstationary
motion of a material object in the aether; \  (2) the model
correctly describes the increase of mass and reduction of
dimensions of a moving object along a path. In this respect the
model complies with the Lorentz hypothesis on the change of the
geometry of a particle in motion. \ (3) The model introduces
proper time, that is Lorentz-invariant, for the description of
behavior of material objects.

      The theory proposed here and in Ref. 14 and here sheds some light
on the hidden dynamics of free moving spinless canonical particle.
The theory is characterized by short-range action and is valid
both for the nonrelativistic and relativistic particle. It is
causal and therefore contains the equation of path [see (15)].
Moreover, this theory give rise to the basic relations of quantum
mechanics [see (35) and (36)], and  its outcome is the
Schr\"odinger wave equation in a relativistic-invariant form [see
(37) and (39)]. Thereby, one can affirm that the contemporary
nonrelativistic quantum theory is really the relativistic one [it
is only necessary that all masses $M_0$ in any system under
consideration are replaced by relativistic masses, $M_0
\rightarrow M_0/\sqrt{1-v_0^2/c^2}$, \ but this is a negligible
correction with accuracy to $v_0^2/c^2$ since $v_0\ll c$].

    Consequently, all these results allow us to regard the vacuum in
the form of an elastic discrete medium and in this case inertons
as elementary excitations of this medium should exist and manifest
themselves experimentally.

\vspace{5mm}
\hspace{4cm} {\bf Acknowledgment}

I am very grateful to Professor Dirk ter Haar for interest in and
support of this work.

\newpage
 {\bf APPENDIX}
\vspace{4mm}

  For the variables $X^s_{(r)}$ and $x^s_{(r)}$ one obtains
from Eq. (5) the equations of extremals:
   $$
{\ddot X}^s_{(r)} + \Gamma^s_{ij}{\dot X}^i_{(r)}{\dot X}^
j_{(r)}  +{\pi\over T_{(r)}} g^{sk} \Bigl\lgroup  {\partial{\hat B}
^{(r)}_{ij}\over \partial X^k_{(r)}} \Bigr\rgroup
\ \ \ \ \ \ \ \ \ \ \ \ \ \ \ \ \ \ \ \ \ \ \ \ \ \ \ \ \ \ \ \
$$
$$
 \ \ \ \ \ \ \ \ \ \ \ \ \ \ \ \ \ \ \ \ \
\times \Bigl( X^i_{(r)}{\dot x}
^j_{(r)} +{{\dot X}^i_{(r)} \vert}_{t_{(r)}=0} x^j_{(r)} \Bigr)
+ {\pi\over T_{(r)}}g^{sk}{\hat B}^{(r)}_{kj}{\dot x}^j_{(r)}=0;
\ \ \ \ \ \ \ \  \eqno (A1)
   $$

   $$
 \ \ \ {\ddot x}^s_{(r)}+{\widetilde\Gamma}^{(r) s}_{ij}{\dot x}^i_{(r)}
 {\dot x}^j_{(r)}
+ {\pi\over T_{(r)}} {\widetilde g}^{(r) sk} \Bigl\lgroup
{\partial {\hat B}^{(r)}_{ij}\over \partial x^k_{(r)}} -
{\partial{\hat B}^{(r)}\over \partial x^j_{(r)}} \Bigr\rgroup
X^i_{(r)}{\dot x}^j_{(r)}   \ \ \ \ \ \ \ \ \ \
   $$
  $$ \ \ \ \ \ \ \ \ \ \ \ \ \ \ \ \ \ \ \ \ \ \ \ \ \ \ \ \ \ \ \ \
\ \ \ \ \ \ -{\pi\over T_{(r)}}{\widetilde g} ^{(r) sk} {\hat
B}^{(r)}_{ki} \Bigl( {\dot X}^i_{(r)} - {{\dot X}^i_{(r)}
\vert}_{t_{(r)} =0} \Bigr) =0; \eqno (A2)
  $$
here, $\Gamma^s_{ij}$ and ${\widetilde\Gamma}^{(r) s}_{ij}$ are
symmetrical connections (see Ref. 19) for the particle and for the
$r$th inerton, respectively; indices $i, j$, and $s$ take the
values 1, 2, 3. The designation $$ {\hat B}^{(r)}_{ij}
\equiv(g_{if}({\hat A}^{-1}{\widetilde g}^{(r)}_{fj})_o )^{1/2}
\eqno (A3) $$ is introduced into (A1) and (A2). The process of
"knocking out" of the next $r$th inerton can be subdivided into
three stages (Ref. 14). At the first stage the interaction
operator in Eqs. (A1) and (A2) is still not engaged, ${\hat
B}^{(r)}_{ij}=0$. Then the termwise difference between these
equations is reduced to the form
  $$
(\ddot X_{(r)}-\ddot x_{(r)}) + (\Gamma^s_{ij}\dot X^i_{(r)}\dot
X^j_{(r)} -{\widetilde\Gamma}^{(r) s}_{ij}\dot x^i_{(r)}\dot
x^j_{(r)}) =0. \ \  \eqno (A4)
  $$
Equation (A4) describes the union of the particle and $r$th
inerton into a common system, and because of this, the
acceleration that one of the partners of the system experiences
coincides with the acceleration that other partner experiences.
Hence the difference in the first set of parentheses in Eq. (A4)
is equal to zero and one gains the relation
  $$
\Gamma^s_{ij} \dot X^i_{(r)} \dot X^j_{(r)}
={\widetilde\Gamma}^{(r) s} _{ij} \dot x^i_{(r)} \dot x^j_{(r)}.
\eqno (A5)
  $$
Coefficients $\Gamma^s_{ij}$ are generated by the particle mass
$M_o$, but coefficients ${\widetilde\Gamma}^{(r) s}_{ij}$ are
generated by the inerton mass $m_{r}$ (see Ref. 14). Hence the
relation $\Gamma^s_{ij}/{\widetilde\Gamma}_{ij}^{(r) s}=
M_o/m_{r}$ holds. If the velocity of the particle at the point of
emission of the $r$th inerton is $v_{or}$ and the initial velocity
of this emitted inerton is $c$, then one obtains
  $$
M_0v^2_{0r} = m_r c^2. \eqno (A6)
  $$
Setting tensors $g_{ij}$ and ${\widetilde g}^{(r)}_{ij}$ are
constant (see Ref. 14), one can reduce Eqs. (A1) and (A2) to the
form when only two terms remain: the first and the last ones, the
latter being transformed as follows:
  $$
g^{sk}{\hat B}^{(r)}_{kj} \rightarrow (m_{r}/M)^{1/2}= v^s_{0r}/c,
\eqno (A7)
  $$
  $$
{\widetilde g}^{(r) sk}{\hat B}^{(r)}_{kj} \rightarrow
(M/m_{r}) ^{1/2} =c/v^s_{0r}. \eqno (A8)
  $$
As the result, Eqs. (A1) and (A2) take the form (one omits the
parentheses for indices $r$):
  $$ \ddot X^s_{r}+{\pi \over
T_{r}}{v_{or}\over c}\dot x^s_{r}=0 \eqno (A9);
  $$
  $$
\ddot x^s_{r}-{\pi\over T_{r}}{c\over v_{or}}
(\dot X^s_{r} - v^s_{or} ) =0. \eqno (A10)
  $$
Initial conditions are chosen in the form
$$
\dot X_r {(t_r + \Delta t_r)\vert}_{t_r=0}=\dot X(\Delta t_r)= v_{0r}
\eqno (A11)
$$
for the particle and
$$
{x^{\perp}\vert}_{t_r=0}=0, \ \ \ \ \ \ \ {{\dot x}^{\perp}\vert}_
{t_r=0}=c \eqno (A12)
$$
for the $r$th inerton.

       The solution of Eqs. (A9) and (A10) is entirely similar to
the solution of the equations of motions carried out
previously$^{(14)}$ in the nonrelativistic limit.

\newpage  {\bf References }

\begin{enumerate}
\item  W. Pauli, Scientia (Milan) {\bf 56}, 65 (1936).

\item  L. de Broglie, Ann. Fond. L. de Broglie {\bf 12}, 399 (1987).

\item   J.S. Bell,\ {\it  Speakable and  Unspeakable  in  Quantum
            Mechanics}  (Cambridge University
            Press, Cambridge,  1987).

\item  A.O. Barut and M. Bozic, Ann. Fond.  L.
           de  Broglie {\bf 15},  67 (1990).

\item   C.L.B. Ryff, Found. Phys, {\bf 20}, 1061 (1990).

\item  M. Yusouff, Preprint IC/83/46, Int. Centre Theor. Phys.
           (Miramare -Trieste, 1983).

\item   P. Multanrzynski and M. Heller, Found.
            Phys. {\bf 20}, 1005 (1990).

\item  C. Wolf, Ann. Fond. L. de Broglie
            {\bf 15}, 189 (1990).

\item  K.E. Plokhotnikov, Doklady Acad. Nauk USSR
           {\bf 316},  332  (1991) (in   Russian).

\item  G. Brightwell and  R.
            Gregory,  Phys.  Rev.  Lett. {\bf 66},  260 (1991).

\item  C.W. Rietdijk, Ann. Fond. L. de Broglie {\bf 16}, 177
            (1991).

\item  P.I. Fomin, in {\it Quantum
             Gravity}. Proc. IV Seminar on  Quantum Gravity, Moscow,
             USSR, 25 - 29 May 1987; Eds. V. Markov,  V.  Berezin, and
             V.P.  Frolov (World Scientific Publishing  Co.,
             Singapore, 1988), p.  813.

\item F. Winterberg, Int. J. Theor. Phys. {\bf 34}, 265 (1995).

\item  V. Krasnoholovets and D. Ivanovsky, Phys.  Essays {\bf 6}, 554
       (1993) \  (also e-preprint xxx.lanl.gov quant/9910023).

\item  B.A. Dubrovin, S.P. Novikov and A.T. Fomenko,
            {\it Modern   Geometry} \break (Nauka,
            Moscow, 1986), p. 24-36  (in Russian).

\item  L.D. Landau and E.M. Lifshits, {\it The Theory of Field}
                 (Nauka, Moscow, 1973), p. 42 (in Russian).

\item B.A. Dubrovin, S.P. Novikov and A.T. Fomenko,
            {\it Modern Geometry} \break (Nauka, Moscow, 1986), p.295
            (in Russian).

\item {\it Ibid.}, p. 291.

\item {\it Ibid.},  pp. 289 - 291.

\item D. ter Haar, {\it Elements of Hamiltonian Mechanics}
            (Nauka, Moscow, 1974), p. 157 (Russian translation).

\item  H. Goldstein, {\it  Classical Mechanics} (Nauka, Moscow,
             1975), p. 306   (Russian translation).

\item L. de Broglie,
            {\it Heisenberg's Uncertainty Relations and the
            Probabilistic Interpretation of Wave Mechanics}
            (Mir,  Moscow, 1986), pp. 42-43 (Russian translation).

\item  E. Fermi, {\it  Notes on Quantum Mechanics} (Mir, Moscow,
             1965), pp. 15-19 (Russian translation).

\item L.I. Schiff, {\it Quantum mechanics} (Izdatelstvo
Inostrannoy Literatury, Mos-\break cow, 1959), p. 33 (Russian translation).

\item {\it Ibid.}, p. 364.

\item  A.A. Sokolov, Yu.M. Loskutov and I.M.
             Ternov, {\it  Quantum  Mechanics}
             (Prosveshchenie, Moscow, 1965), p.55 (in Russian).

\item  L. de Broglie, {\it Heisenberg's Uncertainty Relations and the
            Probabilistic Interpretation of Wave Mechanics}
            (Mir, Moscow, 1986),  pp. 34-42 (Russion translation).

\item B.A. Dubrovin, S.P. Novikov and A.T. Fomenko
            {\it Modern Geometry} \break (Nauka,   Moscow, 1986),
            p.63 (in Russian).

\item  P.G. Bergmann, {\it Introduction to the Theory of
               Relativity}
            \break (Gosudarstvennoe  Izdatelstvo Inostrannoy
               Literatury, Moscow,
            \break 1947),  pp. 72-73, 114-115 (Russian translation).

\item H.A. Lorentz, {\it Aether Theories and Aether Models}
             (Nauchno-
           \break Tekhnicheskoe Izdatelstvo NKTP USSR,
              Moscow - Leningrad, 1936), p. 25  (Russian translation).

\item  {\it Idem}, \  Amst. Proc., {\bf 6}, 809 (1904);
            \  {\bf 12}, 986 (1904).

\item P. Ehrenfest, Zhurn. Russkago Fiz.-Khim.
               Obshchestva
           \break (St.-Petersburg), Part Phys.,
               {\bf 45},  No. 4B, 151 (1913).

\item A. Einstein, Ann. d. Phys., {\bf B 17}, Ser. 4, \   891
            (1905).

\item L.D. Landau and E.M. Lifshits, {\it Hydrodynamics}
                (Nauka, Moscow, 1986), p. 15 (in Russian).

\item {\it Ibid.}, p. 351.
\end{enumerate}

\end{document}